\documentclass[twocolumn,amsmath,amssymb,aps,prl,floatfix,nofootinbib,superscriptaddress]{revtex4-1}

\usepackage{bm} \usepackage{graphicx} 
\usepackage{amsmath}
\usepackage{amsthm,stmaryrd,mathtools}
\usepackage{amssymb} 
\usepackage{epstopdf} 
\usepackage{txfonts}
\usepackage{graphicx}
\usepackage{siunitx}
\usepackage{hyperref}
\hypersetup{hidelinks}
\usepackage{braket}
\usepackage{amsmath}
\usepackage{diagbox}
\usepackage{dcolumn}

\usepackage{graphicx,graphics,times,color}
\usepackage{bm}
\usepackage{longtable}
\usepackage{color}
\usepackage{lipsum}
\usepackage{graphicx}
\usepackage{epstopdf}

\def\be{\begin{equation}}
\def\ee{\end{equation}}
\def\ba{\begin{eqnarray}}
\def\ea{\end{eqnarray}}

\begin{document}

\title{Remote Quantum Sensing with Heisenberg Limited Sensitivity in Many Body Systems}

\author{Gareth Si\^on Jones}
\email{gareth.jones.16@ucl.ac.uk}
\affiliation{Department of Physics and Astronomy University College London, Gower Street, London, WC1E 6BT, U.K.}

\author{Sougato Bose}
\email{s.bose@ucl.ac.uk}
\affiliation{Department of Physics and Astronomy University College London, Gower Street, London, WC1E 6BT, U.K.}

\author{Abolfazl Bayat}
\email{abolfazl.bayat@uestc.edu.cn}
\affiliation{Institute of Fundamental and Frontier Sciences, University of Electronic Science and Technology of China, Chengdu 610051, China}

\date{\today}

\begin{abstract}
Quantum sensors have been shown to be superior to their classical counterparts in terms of resource efficiency. Such sensors have traditionally used the time evolution of special forms of initially entangled states, adaptive measurement basis change, or the ground state of many-body systems tuned to criticality. Here, we propose a different way of doing quantum sensing which exploits the dynamics of a many-body system, initialized in a product state, along with a sequence of projective measurements in a specific basis. The procedure has multiple practical advantages as it: (i) enables remote quantum sensing, protecting a sample from the potentially invasive readout apparatus; and (ii)  simplifies initialization by avoiding complex entangled or critical ground states.  From a fundamental perspective, it harnesses a resource so far unexploited for sensing, namely, the residual information from the unobserved part of the many-body system after the wave-function collapses accompanying the measurements. By increasing the number of measurement sequences, through the means of a Bayesian estimator, precision beyond the standard limit, approaching the Heisenberg bound, is shown to be achievable. 
\end{abstract}

\maketitle

\emph{Introduction.--} 
Quantum sensing is one of the key applications of quantum technologies~\cite{Degen,braun2018quantum}, with various physical realisations including nitrogen vacancies in diamond~\cite{Taylor2008,Nusran,Bonato2015,Said,Waldherr2011,farfurnik2019spin,dolde2011electric,blok2014manipulating,arai2015fourier,holzgrafe2019nanoscale,patel2020sub}, photonic devices~\cite{mitchell2004super,nagata2007beating,taylor2013biological,hou2019control}, ion traps~\cite{leibfried2004toward,biercuk2009optimized,maiwald2009stylus,baumgart2016ultrasensitive,bohnet2016quantum}, cold atoms~\cite{appel2009mesoscopic,leroux2010implementation,louchet2010entanglement,sewell2012magnetic,bohnet2014reduced,hosten2016measurement}, superconducting qubits~\cite{bylander2011noise,bal2012ultrasensitive,yan2013rotating,wang2019heisenberg}, and optomechanical systems~\cite{krause2012high,guzman2014high,bagci2014optical}. The precision of any protocol for sensing an unknown parameter $B$, quantified by the standard deviation $\delta B$, is bounded by the Cram\'er-Rao inequality, i.e. $\delta B \ge 1/\sqrt{M\mathcal{F}}$, where $M$ is the number of samples, and $\mathcal{F}$ is the Fisher Information~\cite{helstrom1969quantum}. For any resource $T$, which can be time~\cite{Cappellaro2012,Nusran,Taylor2008,Waldherr2011,Bonato2015,Said} or number of particles~\cite{Giovannetti2004,Giovannetti2006,Giovannetti2011}, a classical sensor results in $\mathcal{F}{\sim} T$ (the standard limit). In a quantum setup however, by exploiting entanglement, e.g. in the form of a system initialised in a GHZ state~\cite{greenberger1989going}, precision can be dynamically enhanced to $\mathcal{F}{\sim}T^{2}$ (the Heisenberg limit)~\cite{Giovannetti2004, Giovannetti2006, Giovannetti2011}. This enhanced sensitivity persists in the case of open quantum systems too~\cite{beau2017nonlinear,alipour2014quantum}. Since preparing and maintaining GHZ-states is challenging~\cite{kolodynski2013efficient}, alternative approaches, namely exploiting the coherence of a single particle sensor through adaptively updating the measurement basis ~\cite{higgins2007entanglement,Said,Berry2009,Higgins2009,Bonato2015}, and continuous measurements~\cite{gammelmark2014fisher}, have also been shown to exceed the standard limit. However, modifying the basis and continuous measurements may not always be practicable. Therefore, one may wonder whether it is possible to exploit other quantum features, such as projective measurement and its subsequent wave-function collapse, to achieve Heisenberg limited sensitivity?

Many-body systems are resourceful for entanglement in both their ground state~\cite{amico2008entanglement,de2018genuine} and non-equilibrium dynamics~\cite{eisert2015quantum,gogolin2016equilibration,schachenmayer2013entanglement,calabrese2018entanglement,alba2018entanglement,islam2015measuring,ho2017entanglement,bayat2010information,bayat2010entanglement}. Thanks to the enhanced multi-partite entanglement~\cite{guhne2005multipartite,guhne2006energy,giampaolo2013genuine,giampaolo2014genuine,bayat2017scaling,rams2018limits} near criticality, at equilibrium (e.g. in the ground ~\cite{zanardi2008quantum,invernizzi2008optimal,salvatori2014quantum,bina2016dicke,boyajian2016compressed,frerot2017entanglement} or thermal ~\cite{mehboudi2016achieving} state), a strongly-interacting many-body system can be used to sense an external parameter with quantum limited sensitivity. For conventional dynamical strategies with initial entangled states~\cite{Giovannetti2004,Giovannetti2006,Giovannetti2011}, the interactions between particles is often ignored since they cannot enhance precision~\cite{boixo2007generalized,de2013quantum,skotiniotis2015quantum,pang2014quantum,de2013quantum}. However, it would be highly desirable to use the interactions to avoid the complex preparation of entangled states, and use the dynamics to still achieve precision beyond the standard limit.
\begin{figure} \centering
	\includegraphics[width=0.45\textwidth]{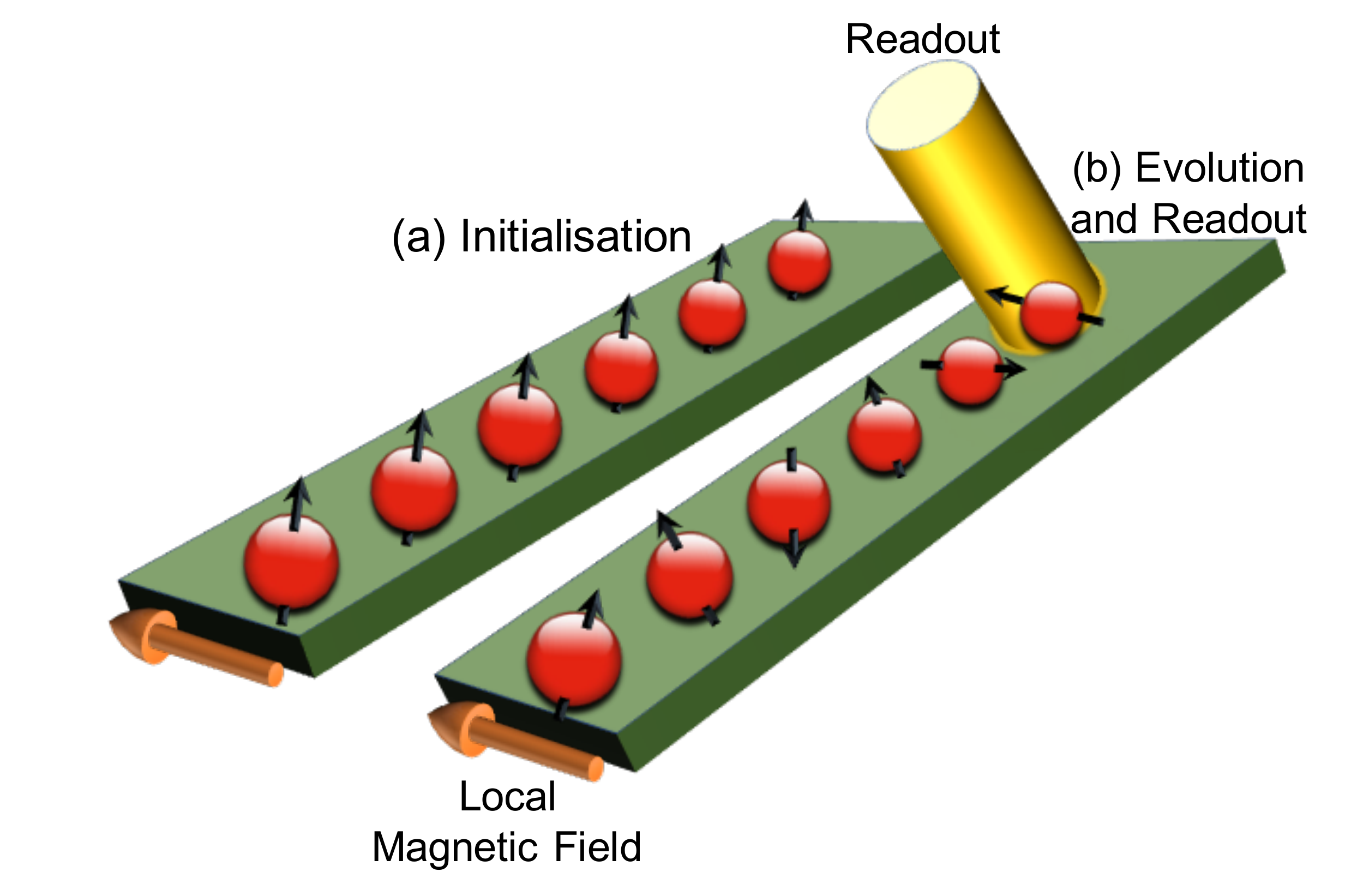}
	\caption{Schematic representation of the remote sensing protocol. (a) A spin chain probe is initialized in a product state. (b) An external magnetic field acting on site 1 drives the dynamics of the whole probe.  The readout is performed sequentially on the last site, separated by intervals of free-evolution.}
	\label{fig:1_Schematic}
\end{figure} 

In this letter, we propose a many-body system, initialized in a product state, as a dynamical probe for sensing a local magnetic field at the first site. A sequence of measurements in a fixed basis, separated by periods of free time evolution, is performed on the last site. With a Bayesian estimator, the local field can be sensed with precision beyond the standard limit, approaching the Heisenberg bound with an increasing number of sequences. This demonstrates that quantum measurement and its subsequent wave-function collapse can harness the information stored in the unobserved part of the many-body system for quantum enhanced sensing. Unlike conventional quantum sensing literature, which often compute a bound using Fisher information, here we have an explicit prescription for obtaining precision beyond the standard limit. 

\emph{Model.--}
We consider a chain of $N$ spin-1/2 interacting particles as a many-body quantum sensor to probe a local magnetic field acting upon the first site, through performing a measurement on the last particle. Without loss of generality, we consider a Heisenberg interaction:
\newline
\begin{equation}
	H = J\sum_{j=1}^{N-1} \bm{\vec{\sigma}}_{j} \cdot \bm{\vec{\sigma}}_{j+1} + B\sigma_{1}^{x}
	\label{eq:1_Hamiltonian}
\end{equation}
where $J$ is the spin exchange coupling, $\bm{\vec{\sigma}}_{j} {=} (\sigma_{j}^{x}, \sigma_{j}^{y}, \sigma_{j}^{z})$ is a vector of the Pauli matrices acting on site $j$, and $B$ is the local magnetic field to be measured, assumed to be in the $x$-direction. The chain is initialized in the ferromagnetic state $\ket{\Psi(0)} {=} \ket{\downarrow \downarrow \downarrow ...}$. In Fig.~\ref{fig:1_Schematic}(a), we present a schematic of the system. In the presence of a local magnetic field $B$, the initial state evolves according to $\ket{\Psi\left(t\right)} {=} e^{-i \hat{H} t} \ket{\Psi_{0}}$. As the system evolves, the quantum state accumulates information about the value of $B$, which can be inferred through a later local measurement in the $z$-direction on site $N$, as depicted in Fig.~\ref{fig:1_Schematic}(b). This provides the distinct advantage of remote quantum sensing, minimizing disturbance of the sample by the measurement apparatus. 
\begin{figure} \centering
	\includegraphics[width=0.45\textwidth]{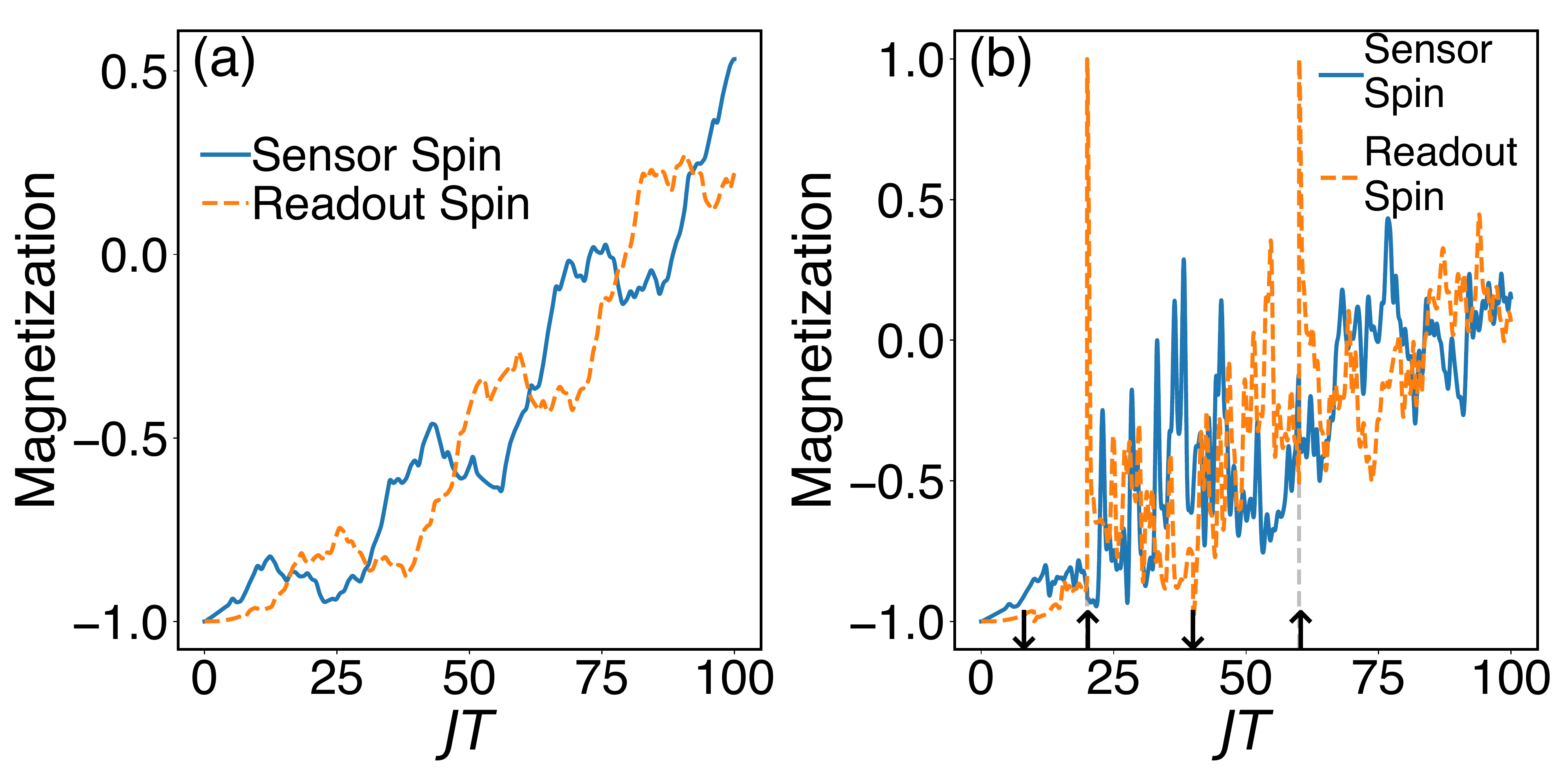}
	\caption{The magnetisation of the first and last sites as a function of time, in a chain of $N{=}10$ with $B/J{=}0.1$. (a) In the absence of measurement, the dynamics of both sites synchronise after an initial transition time. (b) Following a sequence of four measurements, with a sample of outcomes $\left(\downarrow,\uparrow,\downarrow,\uparrow\right)$, the dynamics of both sites are simultaneously affected.}
	\label{fig:2_Magnetisation}
\end{figure} 

In the presence of a non-zero field $B$ the initial state is not an eigenstate of the Hamiltonian, and thus evolves under the action of $H$. By measuring the $N^{\text{th}}$ particle in the $z$-direction, i.e. $\sigma^{z}_{N}$, each measurement outcome appears with the probability $p_{\gamma}{=} \bra{\Psi(t)} \mathcal{M}_{N}^{\gamma} \ket{\Psi(t)}$, where $\mathcal{M}_{N}^{\gamma}{=} \ket{\gamma_{N}}\bra{\gamma_{N}}$ (for $\gamma{=}\uparrow, \downarrow$) is the projection operator for a spin state $\ket{\gamma}$ at site $N$. Therefore, the average magnetization at site $N$ is $m_{N}{=}p_{\uparrow} - p_{\downarrow}$. To see this effect, we look at the magnetization  of both the first and last sites as a function of time in Fig.~\ref{fig:2_Magnetisation}(a) in a system of size $N{=}10$, with $B/J{=}0.1$. As the figure shows, $m_{N}(t)$ evolves in time, roughly synchronizing with the dynamics of $m_{1}(t)$ after a certain delay dictated by the length of the chain. This means that by looking at the dynamics at site $N$, one can estimate the local field $B$.

\emph{Resources for Sensing.--}
The original proposals for quantum enhanced metrology~\cite{Giovannetti2004,Giovannetti2006,Giovannetti2011} took the number of entangled particles in the probe, in the form of a GHZ state, as the key sensing resource. However, the creation and preservation of such states becomes challenging for a large number of particles, making the scheme practically difficult to scale up. Single spin sensors have also been shown to achieve quantum enhanced precision, taking a fixed amount of time as the essential resource~\cite{Taylor2008,Nusran,Cappellaro2012,Said,Bonato2015}. We also consider time as the key resource to quantify the precision of our many-body protocol. While the coherent time evolution of a quantum system is fast, measurement and initialization empirically are one and two orders of magnitude slower respectively~\cite{Bonato2015}. Therefore, for a fixed amount of time, it would be greatly beneficial to reduce the the number of initialisations, and save the time to increase the number of measurements and thus the information about the quantity of interest. This is only possible for a many-body sensor. In this case, entanglement builds up naturally during the evolution and a local measurement results in a partial wave-function collapse. The new state of the system still carries information about the local field, and can be used as the initial state for the next evolution without requiring costly re-initialisation.

\emph{Sequential Measurement Protocol.--} 
In a typical sensing scheme, after each evolution followed by a measurement, the probe is reset, and the procedure is repeated. We call this the standard strategy. Since initialisation is very time consuming, this approach demands a significant time overhead. We propose a profoundly different yet simple strategy to use the time resources more efficiently, exploiting measurement induced dynamics~\cite{burgarth2014exponential,pouyandeh2014measurement,bayat2017scaling,ma2018phase}, and the unique nature of
many-body systems. After initialization, a sequence of $n_{seq}$ successive measurements is performed on the readout spin, each separated by intervals of free evolution, without resetting the probe. The data gathering process is: (i) The system freely evolves as: $\ket{\Psi^{(i)}(\tau_{i})}{=}e^{-iH\tau_{i}} \ket{\Psi^{(i)}(0)}$; (ii) The $i^{th}$ measurement outcome $\ket{\gamma}{=}\ket{\uparrow},\ket{\downarrow}$ on the last site $N$ appears with probability: $p_{\gamma}^{(i)}{=}\bra{\Psi^{(i)}(\tau_{i})} \mathcal{M}_{N}^{\gamma} \ket {\Psi^{(i)}\tau_{i})}$; (iii) As a result of obtaining outcome $\gamma$, the wave-function becomes $\ket{\Psi^{(i+1)}(0)}{=}\frac{\mathcal{M}_{N}^{\gamma} \ket{\Psi^{(i)}(\tau_{i})}}{\sqrt{p^{(i)}_\gamma}}$; (iv) Repeat from 1 until $n_{seq}$ data are gathered. $\ket{\Psi^{(1)}(0)}{=}\ket{\Psi(0)} $ is the probe's ferromagnetic initial state, and $\tau_{i}$ is the evolution time between measurement $i{-}1$ and $i$. After gathering a data sequence of length $n_{seq}$, the probe is reset, and the process repeats to generate a new data sequence. The sequential protocol reduces to the standard case for $n_{seq}{=}1$.

To demonstrate the protocol, in Fig.~\ref{fig:2_Magnetisation}(b) we plot the magnetization $m_{1}$ and $m_{N}$ as a function of time when the system undergoes sequential measurements of $n_{seq}=4$. As the figure shows, with each measurement the magnetization of site $N$ jumps to either $-1$ or $+1$ depending on the measurement outcome. Since the whole state is entangled as a result of the measurement, $m_{1}$ also shows discontinuous jumps in its evolution. The resulting sequence of Fig.~\ref{fig:2_Magnetisation}(b) is $\left(\downarrow,\uparrow,\downarrow,\uparrow\right)$. Due to the entanglement between the readout site and the rest of the system, the generated data in each sequence are highly correlated, which may allow the possibility of harnessing entanglement to surpass the standard limit.

\emph{Bayesian Estimation.--}
In order to infer the magnetic field $B$, the data gathered from the experiment is fed into a Bayesian estimator, which is known to be optimal for achieving the Cram\'er-Rao bound in the limit of large datasets~\cite{Cramer,Helstrom,Holevo,Braunstein1994,Braunstein1996,Paris,Goldstein}. For a sequence of length $n_{seq}$, there are $2^{n_{seq}}$ possible measurement outcomes $\vec{\gamma}{=}(\gamma_{1},\gamma_{1},...,\gamma_{n_{seq}})$, where $\gamma_{k} {=} \uparrow, \downarrow$, obtained at consecutive times while the system has not been reset. By repeating the experiment $M_{sam}$ times, a number of sequences $\left\{\vec{\gamma}\right\}$ will be obtained, which will be used to estimate the magnetic field. By fixing the sequential measurement times $\left\{\tau_{1},...,\tau_{n_{seq}}\right\}$, one can compute $f\left(B|\left\{\vec{\gamma}\right\}\right)$, which is the probability distribution for magnetic field $B$ given a set of measurement outcomes $\left\{\vec{\gamma}\right\}$. Bayes theorem implies:
\begin{equation}
	f\left(B|\left\{\vec{\gamma}\right\}\right) = \frac{f\left(\left\{\vec{\gamma}\right\}|B\right)  f\left(B\right)}{f\left(\left\{\vec{\gamma}\right\}\right)}
	\label{eq:Posterior}
\end{equation}
where $f(B)$ is the prior probability distribution for $B$, $f\left(\left\{\vec{\gamma}\right\}|B\right)$ is the likelihood function, and the denominator $f\left(\left\{\vec{\gamma}\right\}\right)$ is a normalization factor such that the probability distribution sums to $1$. For a given dataset $\left\{\vec{\gamma_{k}}| k{=}1,\cdots,M_{sam}\right\}$, in which each $\vec{\gamma_{k}}$ contains $n_{seq}$ measurement outcomes, the likelihood function is:
\begin{equation}
	f\left(\left\{\vec{\gamma}\right\}|B\right) = \binom{M_{sam}}{k_{1},....,k_{2^{n_{seq}}}} \prod_{j=1}^{2^{n_{seq}}} \left[ f\left(\left\{\vec{\gamma_{j}}\right\}|B\right) \right]^{k_{j}}
	\label{eq:Likelihood}
\end{equation}
where $k_{1} ,\cdots , k_{2^{n_{seq}}}$ represent the number of times that the sequence $\vec{\gamma}_{1}{=}(\uparrow_{1},\uparrow_{2},...,\uparrow_{n_{seq}}) ,\cdots , \vec{\gamma}_{2^{n_{seq}}}{=}(\downarrow_{1},\downarrow_{2},...,\downarrow_{n_{seq}})$ occurs in the whole dataset with the constraint that $k_{1} +\cdots+ k_{2^{n_{seq}}}{=}M_{sam}$, and $\binom{M_{sam}}{k_{1},....,k_{2^{n_{seq}}}}{=}\frac{M_{sam}!}{k_{1}!\cdots k_{2^{n_{seq}}}!}$ is the multinomial operator.

We assume no prior knowledge of the field $B$ is available, and so the prior probability distribution $f(B)$ is uniform over the interval of interest, which without loss of generality is here assumed to be $B/J{\in}|-.2, .2|$. There are several ways to infer $\widehat{B}$ as the estimate for $B$. Here, we take a pessimistic approach, assuming that $\widehat{B}$ is directly sampled from the posterior distribution $f\left(B|\left\{\vec{\gamma}\right\}\right)$. Therefore, the relative error of the estimation is $\left|\widehat{B} - B\right| /B$. Since $\widehat{B}$ is sampled from the probability distribution $f\left(B|\left\{\vec{\gamma}\right\}\right)$, one can quantify the quality of the estimation by defining the dimensionless average squared relative error as:
\begin{equation}
	\delta B^{2} = \int f\left(\widehat{B}|\left\{\vec{\gamma}\right\}\right) \left(\frac{\widehat{B} - B}{B} \right)^{2} d\widehat{B}
	\label{relative_error}
\end{equation}
where the integration is over the interval of interest. A straightforward calculation gives 
\begin{equation}
	\delta B^{2} = \frac{\sigma^{2} + \left(\langle B \rangle - B\right)^{2}}{B^{2}}
	\label{eq:relative_error_solved}
\end{equation}
where $\langle B \rangle$ and $\sigma^{2}$ are respectively the average and variance of the magnetic field with respect to the posterior distribution. Since the variance of the distribution directly appears in $\delta B$, this quantity takes the precision and the variance of the estimation simultaneously.

\emph{Numerical Results.--}
To compare the performance of a standard approach and our sequential protocol, we fix the total execution time. The total time for both strategies can be written as $T_{std}{= }M_{sam}^{std} \left(t_{init} + t_{evo} + t_{meas}\right)$ and $T_{seq}{=}M_{sam}^{seq} \left(t_{init} + n_{seq}t_{evo} + n_{seq}t_{meas}  \right)$, where $t_{init}$, $t_{evo}$, and $t_{meas}$ are the initialization, evolution, and measurement times respectively, and $M_{sam}^{std}$ and $M_{sam}^{seq}$ are the number of samples taken for the standard and sequential protocols. For the sequential algorithm, $t_{evo}$ is taken to be $\left(\tau_{1} + \tau_{2} \cdots \tau_{n_{seq}}\right) / n_{seq}$, the average of all evolution times. Here we take $t_{init} {=} 100  t_{evo}$, and $t_{meas} {=} 10  t_{evo}$ \cite{Bonato2015}. For comparison, it is necessary to take the same total run-time, such that $T_{std}{\approx}T_{seq}$. This relates the number of samples in both strategies as $M_{sam}^{std}{=}\frac{11}{100 + 11n_{seq}} M_{sam}^{seq}$, where the time taken for the initialization, evolution, and measurement have been incorporated. The sequential scheme makes a more efficient use of the key time resource, making $n_{seq}M_{sam}^{seq}$ measurements compared to the standard case, with $M_{sam}^{std}$ measurements.

\begin{figure} \centering
	\includegraphics[width=0.48\textwidth]{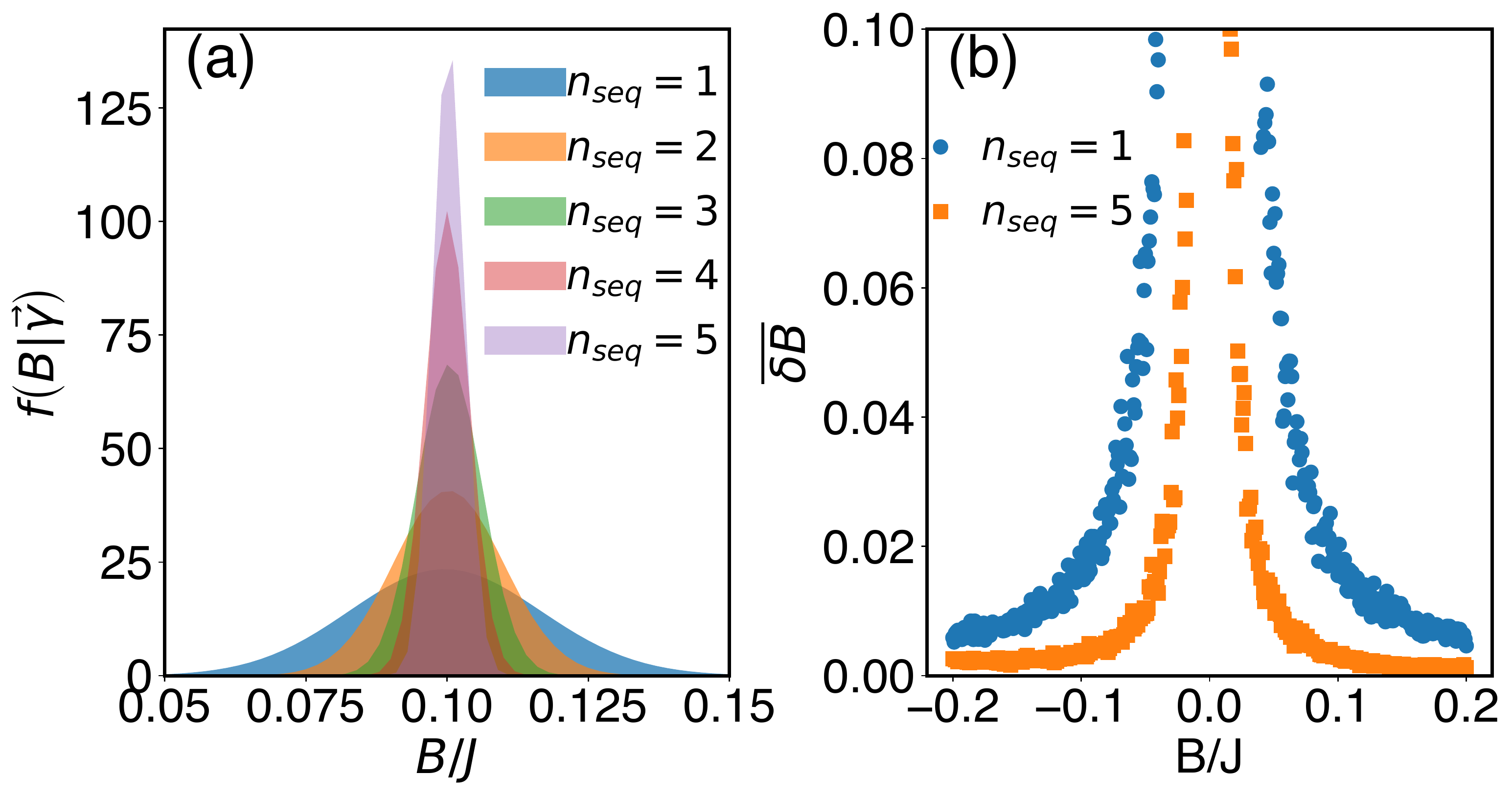}
	\caption{(a) The posterior distribution for a probe of $N{=}5$, with five measurement sequences at times $\tau_{1,\cdots,5} \in \left\{6,8,\cdots,14\right\}$, and with $M_{sam}^{seq}{=}1000$ samples taken at each time, when $B/J{=}0.1$. (b) The average of $\delta B$ from Eq.~(\ref{eq:relative_error_solved}), vs $B/J$, from 100 random samples of a Monte-Carlo simulation.}
	\label{fig:4_PDF_and_Error}
\end{figure} 
The time-evolution of the system is dealt with by exact diagonalization, and the dataset is simulated with a Monte-Carlo approach, in which a measurement outcome is randomly selected from the probability distribution. To show increasing the number of sequences improves precision, in Fig.~\ref{fig:4_PDF_and_Error}(a) we plot the posterior distribution for an increasing number of sequences $n_{seq}$, and for an arbitrarily chosen $B/J{=}0.1$. With each new sequence, the posterior distribution rapidly narrows and thus the variance decreases, providing an increasing precision in the estimate. To assess the performance, we compute $\delta B$ across the whole interval of interest. For each value of $B$, we repeat the protocol $100$ times and take the average error $\overline{\delta B}$. In Fig.~\ref{fig:4_PDF_and_Error}(b), we plot $\overline{\delta B}$ as a function of $B/J$ for both the standard and sequential protocols, with $n_{seq}{=}5$. As $B/J$ tends to zero, the average error diverges, due to the presence of $B$ in the denominator of $\delta B$ in Eq.~\eqref{eq:relative_error_solved}. As $n_{seq}$ is increased, $\overline{\delta B}$ significantly reduces, enhancing the precision. 

\begin{figure} \centering
	\includegraphics[width=0.48\textwidth]{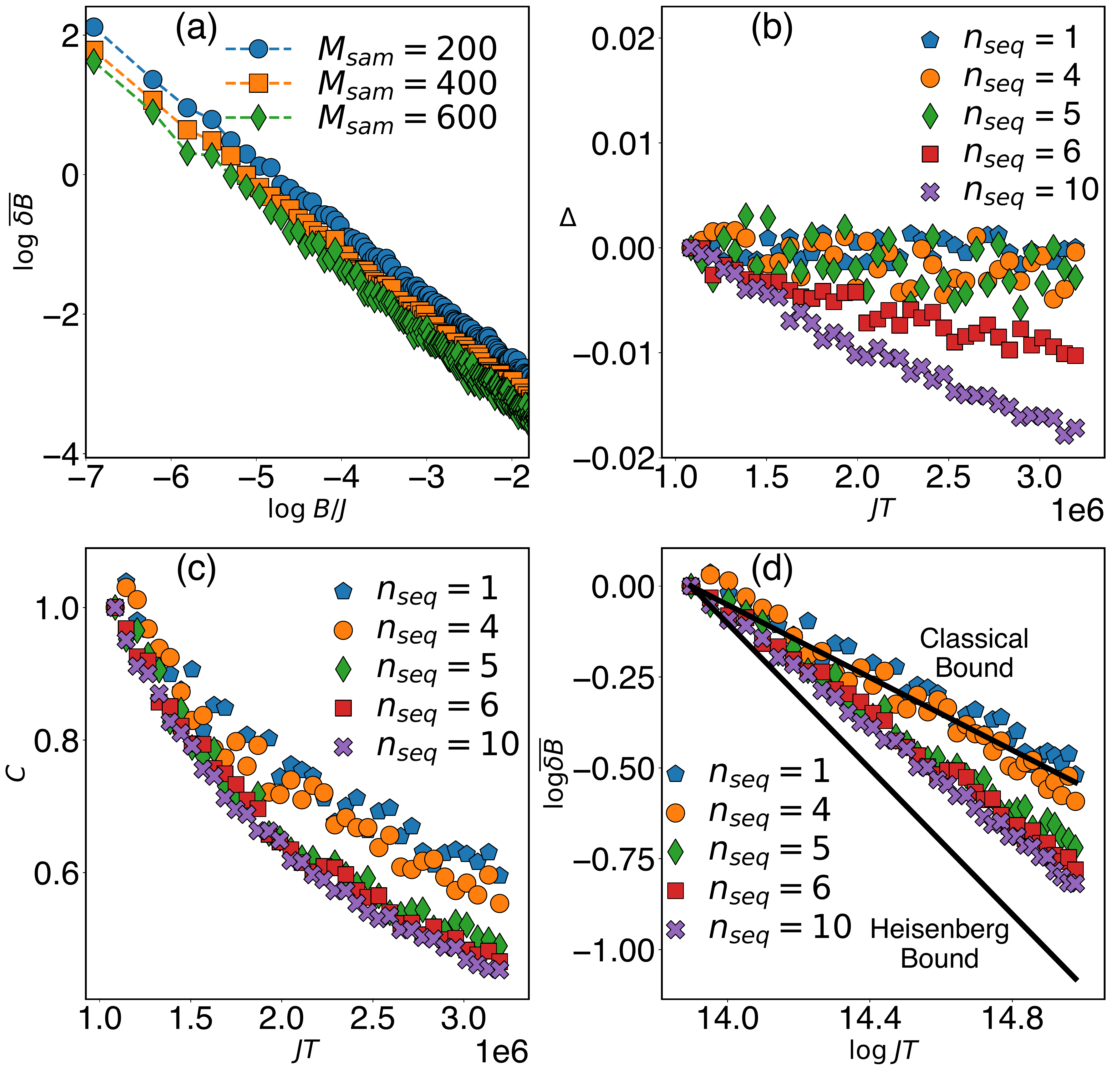}
	\caption{ The performance of the protocol on a system of size $N{=}5$. (a) For $n_{seq}{=}4$, the average error $\overline{\delta B}$ shows an algebraic dependence on $B/J$ for an increasing number of samples $M_{sam}$, and thus total time $T$. (b) The slope  $\Delta$ as a function of $JT$, showing a very weak dependence. (c) The intercept $C$ as vs $JT$, which shows  $C(T){=}AT^{-\alpha}$. (d) The average error $\overline{\delta B}$ versus $JT$ for increasing measurement sequences $n_{seq}$. As $n_{seq}$ increases, the precision surpasses the standard limit, approaching the Heisenberg bound.}
	\label{fig:5_Error_Scaling}
\end{figure}

\emph{Beyond the Standard Limit.--} 
To see the dependence of the sensitivity on the total estimation time, in Fig. \ref{fig:5_Error_Scaling}(a) we plot $\overline{\delta B}$ vs$B/J$ on a log scale (due to the symmetry, we take only values of $B>0$). For a fixed number of sequences and a given total estimation time $T$, the linearity of the curves on the log scale demonstrates that $\overline{\delta B}{=}C(T) B^{\Delta(T)}$, where $\log C(T)$ is the intercept, and $\Delta(T)$ is the slope of each curve in Fig. \ref{fig:5_Error_Scaling}(a). In Fig. \ref{fig:5_Error_Scaling}(b), we plot $\Delta(T)$ vs time for a different number of sequences. Interestingly, $\Delta$ only weakly depends on $T$, which shows that main dependence of $\overline{\delta B}$ on the total time comes from $C(T)$. We can see this dependence explicitly in Fig. \ref{fig:5_Error_Scaling}(c), where increasing time leads to a decrease in $C(T)$, which can be fitted as $C(T){=}A T^{-\alpha}$, where $A$ and $\alpha$ are both constants, independent of time, but dependent on the number of sequences. Remarkably, increasing the number of sequences results in a faster decay of $C(T)$. Since $\Delta(T)$ is almost independent of time, one gets
\begin{equation}
	\overline{\delta B}(T) = AB^{\Delta}T^{-\alpha}
	\label{eq:deltaB_bar}
\end{equation}

This is the main result of this letter, showing precision scaling with respect to the total time $T$. Without loss of generality, for a fixed value of $B/J{=}0.1$, in Fig.~\ref{fig:5_Error_Scaling}(d), we plot $\overline{\delta B}$ as a function of time for various values of $n_{seq}$. Increasing the number of sequences results in a sensitivity scaling beyond the standard limit, approaching the Heisenberg bound. In Table \ref{tab:table_of_alpha}, we summarize the values of $\alpha$ for two different values of $B$ and an increasing number of sequences, clearly showing the transition from classical to quantum limited scaling. 

\begin{table}[h]
	\centering
	\caption{Values of the scaling constant $\alpha$ are found for different values of $B/J$, and an increasing number of sequences.}
	\begin{tabular}{|l||*{5}{c|}}\hline
		\backslashbox{$\alpha$}{$n_{seq}$}
		&\makebox[3em]{1}&\makebox[3em]{4}&\makebox[3em]{5}
		&\makebox[3em]{6}&\makebox[3em]{10}\\\hline\hline
		for $B/J{=}0.1$ & 0.490 & 0.565 & 0.680 & 0.731 & 0.770\\\hline
		for $B/J{=}0.2$ & 0.491 & 0.562 & 0.677 & 0.725 & 0.758\\\hline
	\end{tabular}
	\label{tab:table_of_alpha}
\end{table}

\emph{Conclusions.--}
In this letter, we have proposed a new strategy for sensing beyond the standard quantum limit using many-body dynamics, without requiring a prior-entangled initial state, adaptive measurement basis change, or tuning a system to quantum criticality. Starting from a pure product state, our protocol employs a sequence of measurements in a single basis, the accompanying wave-function collapse, and the leftover information in the unobserved part of the system, so that the total sensing time can be used more efficiently. This may be a natural way of sensing, applicable to a wide variety of physical systems, as we simply exploit the inherent interactions in a system as well as quantum measurements. A corollary of the  many-body nature is that it enables remote sensing, where the synchronization of the dynamics between the two ends of a spin chain plays a crucial role. This is highly beneficial for sensitive systems where the measurement apparatus can destructively affect the sample of interest.

\emph{Acknowledgements.}--
AB acknowledges the National Key R\&D Program of China, Grant No. 2018YFA0306703. GSJ acknowledges Engineering and Physical Sciences Research Council (Grant No. EP/L015242/1). SB thanks EPSRC grant EP/R029075/1 (Non-Ergodic Quantum Manipulation). 


%

\end{document}